\documentclass[aps,prl,superscriptaddress,twoside,showpacs,
nofootinbib,10pt,floatfix,twocolumn,showpacs]{revtex4-1}
\usepackage{graphicx}
\usepackage{mathrsfs}
\usepackage{amsmath,bbm}
\usepackage{CJK}
\usepackage{tikz}
\usepackage[colorlinks,hyperindex,citecolor=black,
	linkcolor=black,urlcolor=black]{hyperref}
\usepackage{enumerate}

\newcommand\<{\langle}
\renewcommand\>{\rangle}
\renewcommand\d{\partial}
\newcommand\tr{\mathop{\mathrm{Tr}}}

\begin{document}

\begin{CJK*}{UTF8}{gbsn}
\title{Magnetic field dependence of Delta isobars properties in a Skyrme model}
\author{Bing-Ran He (何秉然)}
\email[E-mail: ]{hebingran@njnu.edu.cn}
\affiliation{
Department of Physics, Nanjing Normal University, Nanjing 210023, P.R. China
}

\date{\today}

\begin{abstract}
The properties of $\Delta$ isobars in a uniform magnetic field are investigated. In the weak magnetic field region, the general relations between magnetic moment of nucleons and $\Delta$ isobars are given. 
In the strong magnetic field region, the mass and size of $\Delta$ isobars depend on the increasing of magnetic field strength in different ways: the effective  mass of $\Delta^{++}$, $\Delta^{+}$ and $\Delta^{0}$ first decreases and then increases, consequently, the size of $\Delta^{++}$, $\Delta^{+}$ and $\Delta^{0}$ first increases and then decreases; whereas, the effective mass of $\Delta^{-}$ always increases, and consequently, the size of $\Delta^{-}$ always decreases. 
The estimation shows in the core part of the magnetar, the equation of state for $\Delta$ isobars depends on the magnetic field, which affects the mass limit of the magnetar.

\end{abstract}


\pacs{12.39.Dc, 12.39.Fe, 11.30.Rd, 13.40.Em}

\maketitle
\end{CJK*}
\emph{Introduction. ---}
The strong magnetic field exists widely in heavy ion collider and magnetars~\cite{Kharzeev:2012ph,Miransky:2015ava,Skokov:2009qp}.  
The Skyrme model study of proton and neutron in a uniform magnetic field shows that, the mass and shape of proton and neutron depend on the strength of magnetic field~\cite{He:2016oqk}, thus the equation of state for magnetar is modified. 

The $\Delta$ isobars play very important role in nuclear physics~\cite{Cattapan:2002rx}. For example, $\Delta$ baryon media exists in the heavy ion collider~\cite{Typel:1999yq}  and neutron stars~\cite{Kolomeitsev:2016ptu,Li:2018qaw}. 
Science $\Delta$ isobars have internal structures, in the strong magnetic field, the properties of $\Delta$ baryon could be changed by the magnetic field dramatically.  
The magnetic response of $\Delta$ baryons could cause observable consequences in the heavy ion collision and magnetar. 

Another mystery of the $\Delta$ isobars is the wide range of the magnetic moment given by experimental results~\cite{Patrignani:2016xqp,Tanabashi:2018oca}. Great efforts had been made to predict the magnetic moment of $\Delta$ isobar states, but still leave us lot of challenges
~\cite{Pascalutsa:2006up,Aubin:2008qp,Slaughter:2011xs,Kaur:2015fdm}. 

In this letter, $\Delta$ isobars in a uniform magnetic field are studied in the semi-classical quantization approach of Skyrme model~\cite{Skyrme:1962vh,Adkins:1983ya}. 
Because the wave functions for $\Delta$ isobars are different, the magnetic response of $\Delta$ isobars are different.  
In the weak magnetic field region, the general relations between magnetic moment of nucleons and $\Delta$ isobars are given. By using the experimental results of $\mu_p$ and $\mu_n$, the magnetic moments of $\Delta$ isobars are given, which are consistent with the experimental results. 
In the strong magnetic field region, it is found that with the increasing of the magnetic field strength, the effective  mass of $\Delta^{++}$, $\Delta^{+}$ and $\Delta^{0}$ first decreases and then increases, consequently, the size of $\Delta^{++}$, $\Delta^{+}$ and $\Delta^{0}$ first increases and then decreases. On the other hand, the effective mass of $\Delta^{-}$ always increases, and consequently, the size of $\Delta^{-}$ always decreases.  
Finally, since both the mass and size of $\Delta$ isobars depend on the strength of the magnetic field, the equation of state for $\Delta$ isobars are influenced by the magnetic field, which could affect the properties of magnetar. 

\emph{The model. ---}
The minimal action of the model for the present propose contains two parts:
\begin{equation}
\Gamma=\int d^4 x \mathscr{L}+\Gamma_{\rm WZW}\,,
\label{total_action}
\end{equation}
where $\mathscr{L}$ represents the pion dynamics which is expressed as
\begin{eqnarray}
\mathscr{L}&=&\frac{f_\pi^2}{16}\tr(D_\mu U ^\dag D^\mu U) + \frac{1}{32g^2}\tr([U ^\dag D_\mu U, U^\dag D_\nu U]^2)\nonumber\\
&&+\frac{m_\pi^2 f_\pi^2}{16}\tr(U+U^\dag-2)
\,.\label{lagrangian}
\end{eqnarray}
Here $g$ is a dimensionless coupling constant, and $f_\pi$ and $m_\pi$ are the decay constant and the mass of pion, respectively. 
The covariant derivative for $U$ is defined as 
 $D_\mu U = \partial_\mu U - i \mathcal{L}_\mu U + i U \mathcal{R}_\mu$,  
where 
$\mathcal{L}_\mu=\mathcal{R}_\mu= e Q_{\rm B} {\mathcal{V}_B}_\mu + e Q_{\rm E} H_\mu$ for the present purpose. 
Here $e$ is the unit electric charge, $Q_{\rm B}=\frac{1}{3}\mathbbm{1}$ is the baryon number charge matrix, $Q_{\rm E}=\frac{1}{6}\mathbbm{1}+\frac{1}{2}\tau_3$ is the electric charge matrix,  $\mathbbm{1}$ is the rank $2$ unit matrix, and $\tau_3$ is the third Pauli matrix. 
The external gauge field ${\mathcal{V}_B}_\mu$ is corresponding to the $U(1)_\mathcal{V}$ baryon number,
In the symmetric gauge, the magnetic field $H_\mu$ is expressed as $
H_\mu  = - \frac{1}{2} B  y \eta_\mu^{\;\;1} + \frac{1}{2} B  x \eta_\mu^{\;\;2}
$, 
where $\eta$ is the geometry with ${\rm diag}(+1,-1,-1,-1)$. 

The last part in action \eqref{total_action}, i.e. $\Gamma_{\rm WZW}\equiv\int d^4x \mathscr{L}_{\rm WZW}$  represents the chiral anomaly effects, which is given in Refs.~\cite{Wess:1971yu,Witten:1983tw}.

Following Refs.~\cite{Holzwarth:1985rb,He:2015zca}, in the elliptic coordinate system, $x$, $y$, and $z$ in are expressed as 
\begin{eqnarray}
x&=&c_\rho r \sin (\theta )\cos (\varphi ) \,,\nonumber\\
y&=&c_\rho r \sin (\theta )\sin (\varphi ) \,,\nonumber\\
z&=&c_z r \cos (\theta )\,,\label{xyzansatzs}
\end{eqnarray}
where $c_\rho$ and $c_z$ are positive dimensionless parameters, 
$r\equiv \sqrt{\frac{x^2}{c_\rho^2} + \frac{y^2}{c_\rho^2} + \frac{z^2}{c_z^2}}$, and $\theta$ and $\varphi$ are polar angles with $\theta\in[0,\pi]$ and $\varphi\in[0,2\pi]$. 
In the Cartesian coordinate system, $U$ is expressed as 
\begin{eqnarray}
U=\cos(F(r))\mathbbm{1} + \frac{i \sin(F(r))}{r}\Big(  \frac{\tau_1}{c_\rho}x + \frac{\tau_2}{c_\rho}y +  \frac{\tau_3}{c_z}z\Big)\,.\label{uansatzs}
\end{eqnarray}

In the semi-classical quantization approach~\cite{Adkins:1983ya,Braaten:1988cc,Krusch:2002by}, the spin and iso-spin are obtained by rotating the ansatz equations \eqref{xyzansatzs} and \eqref{uansatzs} in both spatial space and iso-spin space, i.e. 
$\hat{U} = A(U (R))A^\dag$, 
where $A$ and $R$ are the rotation matrix of isospin space and spatial space in $x-y$ plane, respectively. Here $A$ and $R$ satisfy that
$A^{-1} \dot A = \frac{i}{2} \omega_a \tau_a$ and 
$(R^{-1} \dot R)_{ij} = -\epsilon_{ij3} \Omega_3 $, where $a=1,2,3$ and $i,j=1,2$. 

By replacing $U$ to $\hat{U}$ in action \eqref{total_action}, a new action is obtained as $\hat\Gamma = \int d^4x (\hat {\mathscr{L}} + \hat {\mathscr{L}}_{\rm WZW}) = \int d^4x \hat {\mathscr{L}}_{\rm total}$. 
By taking a functional derivative of the action with $\omega_a$ and $\Omega_3$, the canonical conjugate momenta of the isospin and spin are obtained  as 
$I_a
=\left.\frac{\d \hat {\mathscr{L}}_{\rm total}
}{\d\omega_a}\right|_{{\mathcal{V}_B}_\mu\to0}
$ and
$J_3
=\left.\frac{\d \hat {\mathscr{L}}_{\rm total}
}{\d\Omega_3}\right|_{{\mathcal{V}_B}_\mu\to0}
$, 
respectively.

\emph{The general relations between nucleons and $\Delta$ isobars magnetic moment in the week magnetic region. ---}
For the present analysis, the $N_C$ counting are $f_\pi\sim\mathcal{O}(N_C^{1/2})$, $g\sim\mathcal{O}(N_C^{-1/2})$, $m_\pi\sim\mathcal{O}(N_C^{0})$, $eB\sim\mathcal{O}(N_C^{0})$, $\omega_a\sim\mathcal{O}(N_C^{-1})$ and $\Omega_3\sim\mathcal{O}(N_C^{-1})$. 
Thus, up to $\mathcal{O}(N_C^{-1})$, the Hamiltonian is obtained as 
\begin{eqnarray}
{\mathcal{H}}= \sum_{a=1,2,3} \left(\omega_a I_a\right) + \Omega_3 J_3 - \left.\hat{\mathscr{L}}_{\rm total}\right|_{{\mathcal{V}_B}_\mu\to0}\,. 
\end{eqnarray}

The baryon mass and the baryon magnetic moment are obtained by $M_{\Psi}= \<{\Psi}|\int dV \mathcal{H}|{\Psi}\>$ and $\mu_{\Psi}=-\frac{\d M_{\Psi}}{\d (e B)}$, respectively. 
Here $|{\Psi}\>$ expresses the wave functions for proton, neutron, $\Delta^{++}$, $\Delta^{+}$, $\Delta^{0}$ and $\Delta^{-}$ which are given in Refs.~\cite{Adkins:1983ya,Wong:2002eh}. 

\begin{table}[hbt]
\centering
\caption{The general relations between nucleon and $\Delta$ isobars magnetic moment ($|eB| \to 0$)}
\label{tab:mu_deltas}
\begin{tabular}{ p{1cm} p{3cm} p{3cm} }
\hline\hline
 & $J_3=3/2$ & $J_3=1/2$\\
\hline
$\mu_{\Delta^{++}}$ & $\frac{3}{5} \left(4 \mu _p+\mu _n\right)+3\mu_I$ & $\frac{1}{5} \left(4 \mu _p+\mu _n\right)+3\mu_I$ \\
$\mu_{\Delta^{+}}$ & $\frac{3}{5} \left(3 \mu _p+2 \mu _n\right)+\mu_I$ & $\frac{1}{5} \left(3 \mu _p+2 \mu _n\right)+\mu_I$ \\
$\mu_{\Delta^{0}}$ & $\frac{3}{5} \left(2 \mu _p+3 \mu _n\right)-\mu_I$ & $\frac{1}{5} \left(2 \mu _p+3 \mu _n\right)-\mu_I$ \\
$\mu_{\Delta^{-}}$ & $\frac{3}{5} \left(\mu _p+4 \mu _n\right)-3\mu_I$ & $\frac{1}{5} \left(\mu _p+4 \mu _n\right)-3\mu_I$ \\
\hline\hline
\end{tabular}
\end{table}

In an extreme week magnetic field, magnetic moments of $\Delta$ isobars are written in nucleons as shown in Table~\ref{tab:mu_deltas}. 
Table~\ref{tab:mu_deltas} shows that, the magnetic moment of a $\Delta$ isobar state is constructed by two parts, one part is related to the strength of spin and another part is related to the strength of iso-spin. 
The magnetic moment relates to spin part is a combination of proton and neutron magnetic moment $\mu_p$ and $\mu_n$. 
The magnetic moment relates to iso-spin part $\mu_I$ can be determined numerically as about $\mu_I=-0.045\,[\mu_N]$, which is much smaller than the magnitude of $\mu_p$ and $\mu_n$. 

By taking experimental value of proton and neutron magnetic moment as input, the theoretical value and experimental value of $\Delta$ isobars magnetic moments ($|eB| \to 0$) are shown in  Table~\ref{tab:mu_skyrmion_exp}.
\begin{table}[hbt]
\centering
\caption{The $\Delta$ magnetic moment ($|eB| \to 0$) by taking experimental value of proton and neutron magnetic moment $\mu_p=2.793\,[\mu_N]$ and $\mu_n=-1.913\,[\mu_N]$ as input ~\cite{Tanabashi:2018oca} 
}
\label{tab:mu_skyrmion_exp}
\begin{tabular}{ l  c  c  c }
\hline\hline
 & \hspace{0.3cm}$J_3=3/2$ \hspace{0.3cm}& \hspace{0.3cm} $J_3=1/2$ \hspace{0.3cm}& \hspace{0.3cm} Exp.~\cite{Patrignani:2016xqp}  \hspace{0.3cm}\\
\hline
$\mu_{\Delta^{++}}$ & $5.42$ & $1.72$ & $5.6\pm1.9$ \\
$\mu_{\Delta^{+}}$ & $2.69$ & $0.87$ & $2.7^{+1.0}_{-1.3}\pm1.5\pm3$ \\
$\mu_{\Delta^{0}}$ & $-0.05$ & $0.01$ &   \\
$\mu_{\Delta^{-}}$ & $-2.78$ & $-0.84$ &  \\
\hline\hline
\end{tabular}
\end{table}
Table~\ref{tab:mu_skyrmion_exp} shows that the theoretical predictions of  $\mu_{\Delta^{++},J_3=3/2}$ and $\mu_{\Delta^{+},J_3=3/2}$ are consistent with the experimental results~\cite{Patrignani:2016xqp}. 

By cancelling the effects of iso-spin shown in Table~\ref{tab:mu_deltas}, one gets the relation 
as  $\mu_{\Delta^{++},J_3={3/2}} + \mu_{\Delta^{-},J_3={3/2}} :\mu_{\Delta^{++},J_3={1/2}} + \mu_{\Delta^{-},J_3={1/2}} =  \mu_{\Delta^{+},J_3={3/2}} + \mu_{\Delta^{0},J_3={3/2}} : \mu_{\Delta^{+},J_3={1/2}} + \mu_{\Delta^{0},J_3={1/2}}
=3:1$ and  $3\mu_{\Delta^{+},J_3={3/2}} - \mu_{\Delta^{++},J_3={3/2}}
= 3\mu_{\Delta^{0},J_3={3/2}} - \mu_{\Delta^{-},J_3={3/2}}
= \mu_{\Delta^{++},J_3={3/2}} + \mu_{\Delta^{-},J_3={3/2}}
= \mu_{\Delta^{+},J_3={3/2}} + \mu_{\Delta^{0},J_3={3/2}}
= (\mu_{\Delta^{++},J_3={3/2}} + 3\mu_{\Delta^{0},J_3={3/2}})/2
= (3\mu_{\Delta^{+},J_3={3/2}} + \mu_{\Delta^{-},J_3={3/2}})/2
= 3(\mu_p+\mu_n)$. 
At present, there are only limited and rough results available for the magnetic moment of $\Delta$ isobars, by taking the central value of experimental results shown in Table~\ref{tab:mu_skyrmion_exp}, i.e. $\mu_{\Delta^{++},J_3={3/2}}=5.6\,[\mu_N]$ and $\mu_{\Delta^{+},J_3={3/2}}=2.7\,[\mu_N]$, one can check the experimental results satisfy the relation $3\mu_{\Delta^{+},J_3={3/2}}-\mu_{\Delta^{++},J_3={3/2}}:
\mu_p+\mu_n=3:1.056\simeq3:1$, which is consistent with Table~\ref{tab:mu_deltas}.

\emph{Numerical results. ---} 
The equation of motion for $\Delta$ isobars are obtained from $\<{\Psi}|\hat\Gamma|{\Psi}\>$ at $\mathcal{O}(N_C)$ order, respectively. 
A standard set of parameters in $N_B=1$ sector is considered as 
$m_\pi=138$ [MeV], $f_\pi=108$ [MeV], and $g=4.84$~\cite{Adkins:1983ya,Adkins:1983hy}. 

\begin{figure}[htb]
\centering
\includegraphics[scale=0.46]{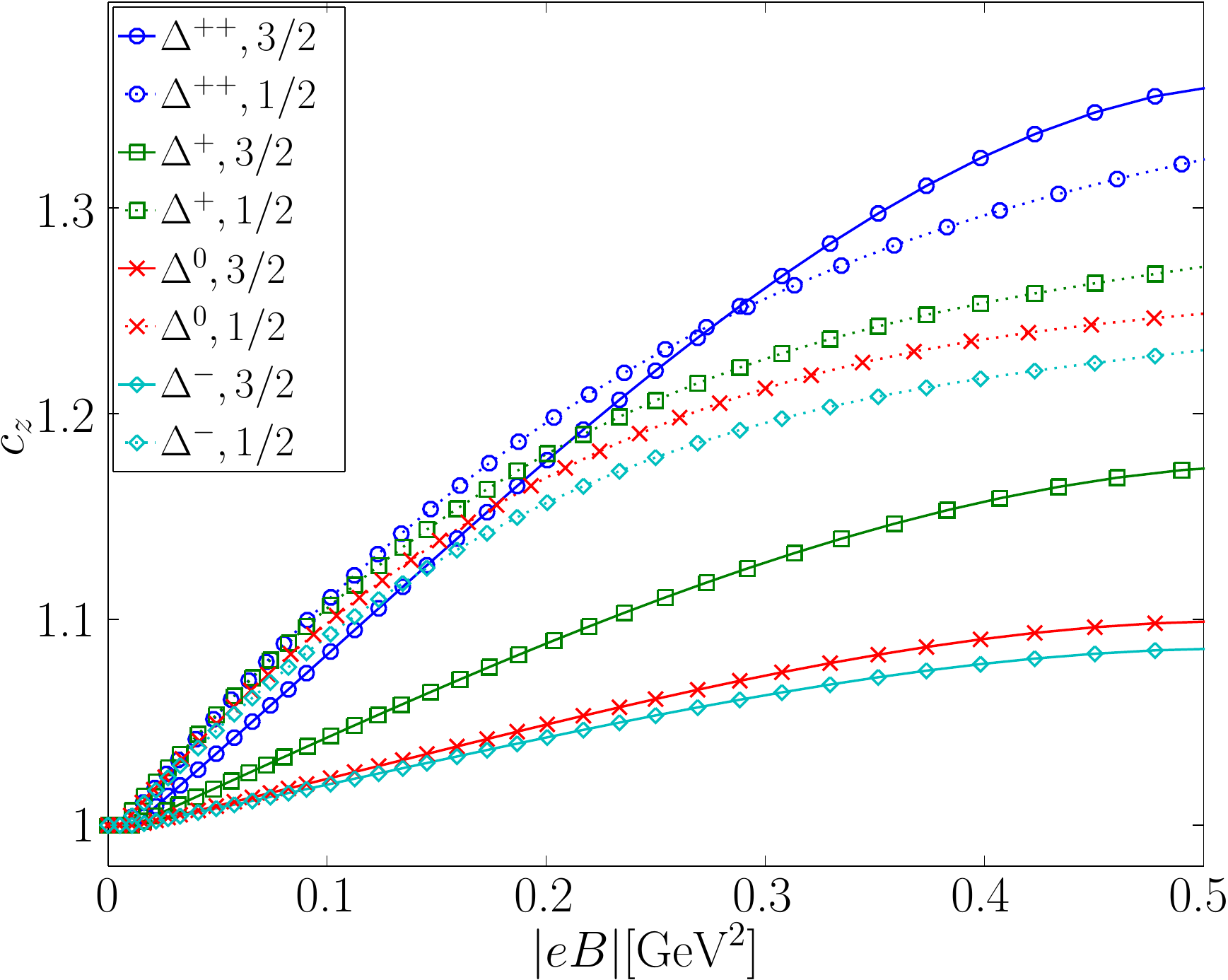} 
\caption{$|e B|$ dependence of $c_z$ for $\Delta^{++}$, $\Delta^{+}$, $\Delta^{0}$ and $\Delta^{-}$.
}
\label{fig:bmag_cz}
\end{figure}

With no loss of generality, $c_\rho$ is determined as $c_\rho= 1/\sqrt{c_z}$~\cite{He:2015zca}.  
For a $\Delta$ isobar state of a given $|e B|$, the parameter ${c}_z$ is fixed to minimize the corresponding $\Delta$ mass. 
The $|e B|$ dependence of ${c}_z$ for $\Delta$ isobar states are shown in Fig.~\ref{fig:bmag_cz}. 
In Fig.~\ref{fig:bmag_cz}, one finds that a stronger magnetic field corresponding to a bigger $c_z$, the reason is that a stronger magnetic field generates more restriction force of charged meson $\pi^{+,-}$ in the $x-y$ plane, which makes the shape of $\Delta$ isobars more stretched along $z$-axis.

\begin{figure}[htb]
\centering
\includegraphics[scale=0.46]{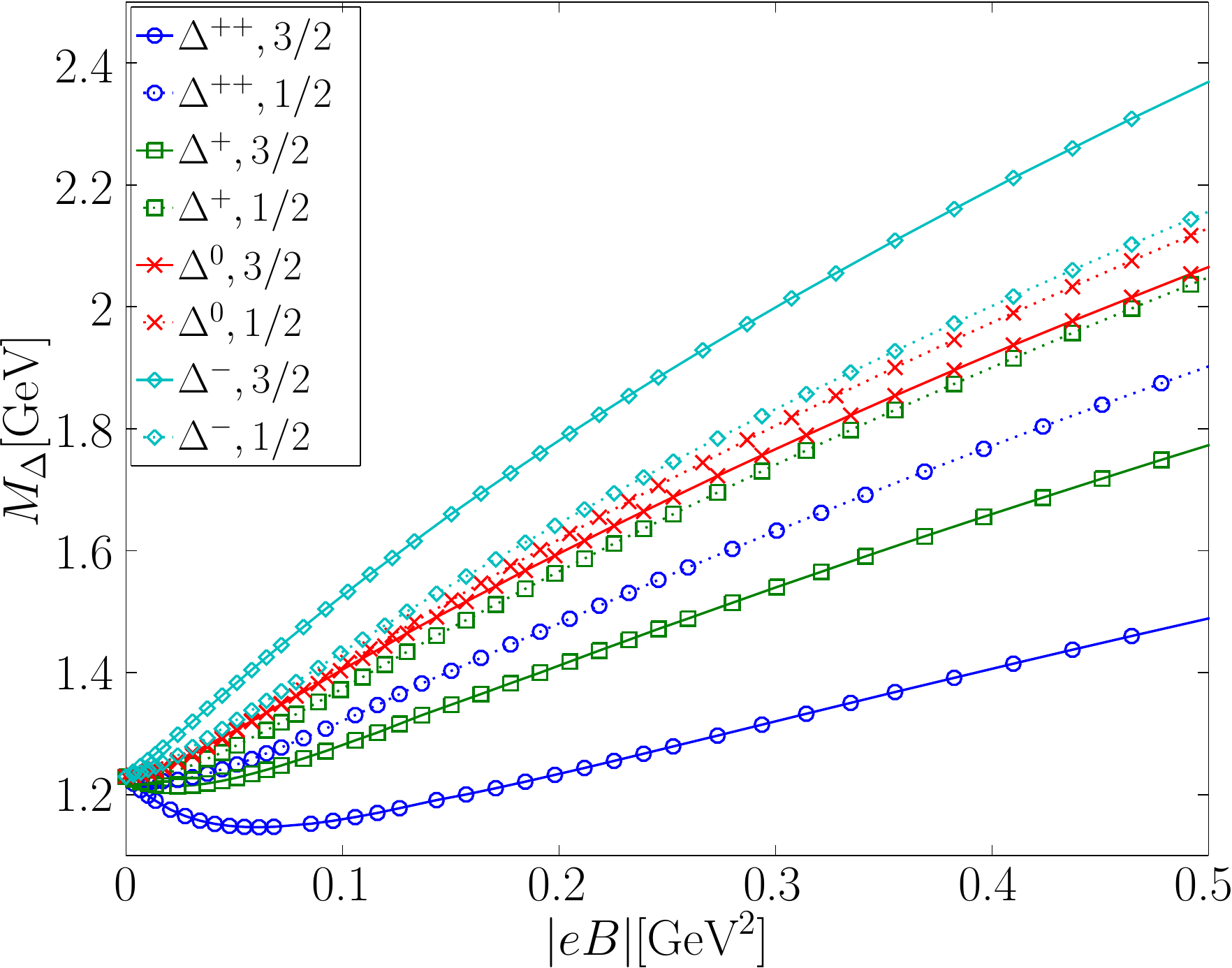} 
\caption{$|e B|$ dependence of $M_{\Delta^{++}}$, $M_{\Delta^{+}}$, $M_{\Delta^{0}}$ and $M_{\Delta^{-}}$.
}
\label{fig:bmag_mpmn}
\end{figure}

The $|e B|$ dependence of the mass of $\Delta$ isobar states are shown in Fig.~\ref{fig:bmag_mpmn}. 
In Fig.~\ref{fig:bmag_mpmn}, the curve shows that with increasing the strength of magnetic field, the mass of $\Delta^{++}$, $\Delta^{+}$ and $\Delta^{0}$ first decreases then increases, whereas $\Delta^{-}$ always increases. 
The reason is that the Hamiltonian of $\Delta$ isobar states contains linear term of $(eB)$ and higher order terms of $(eB)$. 
The sign of the linear term of $(eB)$ for $\Delta^{++}$, $\Delta^{+}$ and $\Delta^{0}$ states is different from $\Delta^{-}$ state, which makes their mass decrease and increase when the magnetic field is weak, respectively. 
The higher order terms of $(eB)$ always increase the mass of $\Delta$ isobars, which makes their mass increase when the magnetic field is strong. 

\begin{figure}[htb]
\centering
\includegraphics[scale=0.46]{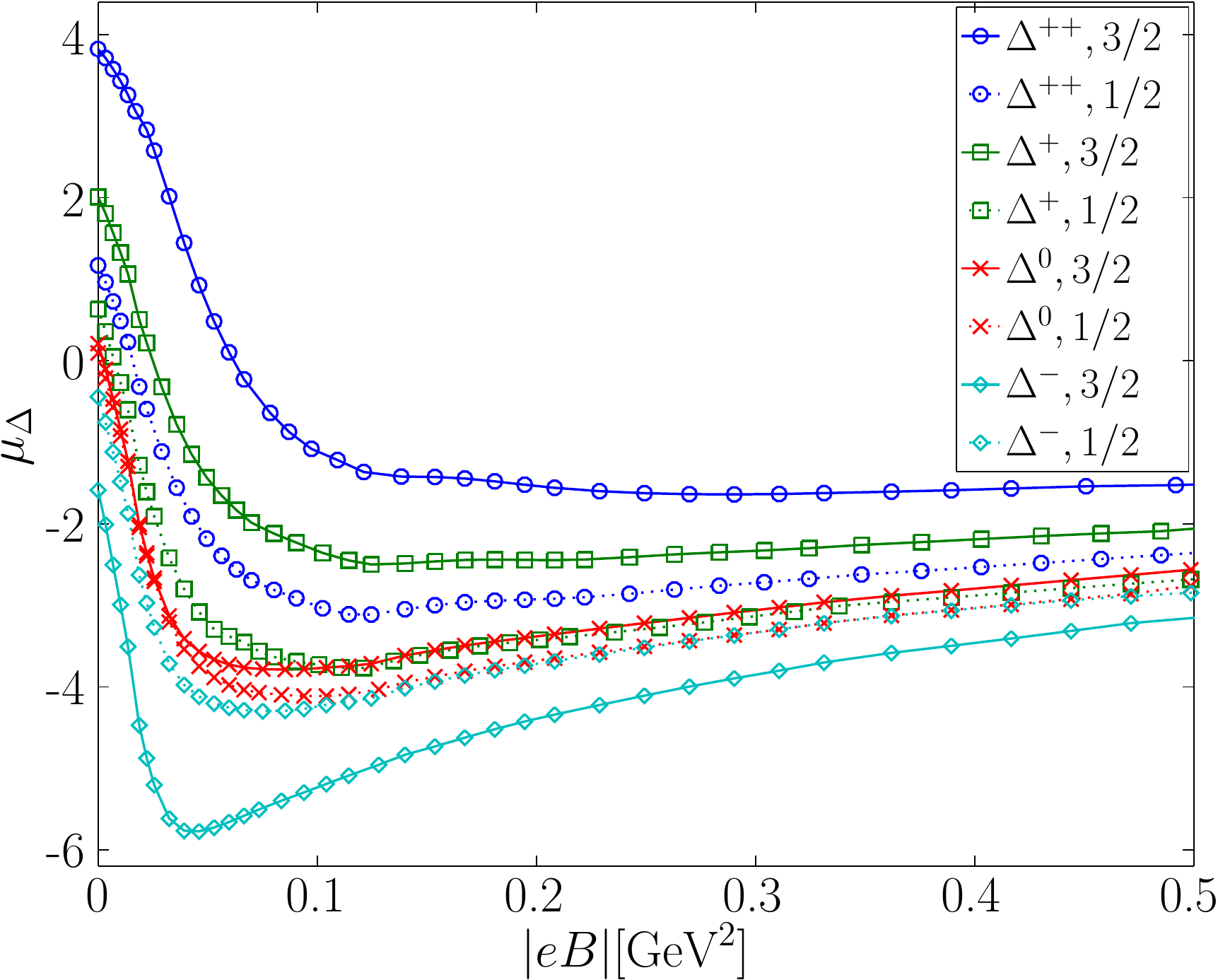} 
\caption{$|e B|$ dependence of $\mu_{\Delta^{++}}$, $\mu_{\Delta^{+}}$, $\mu_{\Delta^{0}}$ and $\mu_{\Delta^{-}}$. 
}
\label{fig:bmag_mupmun}
\end{figure}

The $|e B|$ dependence of $\mu_{\Delta^{++}}$, $\mu_{\Delta^{+}}$, $\mu_{\Delta^{0}}$ and $\mu_{\Delta^{-}}$ are shown in Fig.~\ref{fig:bmag_mupmun}. 
Fig.~\ref{fig:bmag_mupmun} shows that with increasing the strength of magnetic field, the magnitude of $\mu_{\Delta^{++}}$, $\mu_{\Delta^{+}}$ and $\mu_{\Delta^{0}}$ first decreases and then increases, whereas the magnitude of $\mu_{\Delta^{-}}$ first increases and then decreases. 
Notice that for an extreme weak magnetic field $|eB|\sim 0$, the magnetic moment of proton and neutron in the Skyrme model are $\mu_p=1.94\,[\mu_N]$ and $\mu_n=-1.21\,[\mu_N]$~\cite{Adkins:1983hy,He:2016oqk}. 
By using the general relations between nucleons and $\Delta$ isobars magnetic moment when $|eB| \to 0$, which is shown in Table~\ref{tab:mu_deltas}, one can easily get that for $J_3=3/2$ states $\mu_{\Delta^{++}}=3.80\,[\mu_N]$, $\mu_{\Delta^{+}}=2.00\,[\mu_N]$, $\mu_{\Delta^{0}}=0.20\,[\mu_N]$, $\mu_{\Delta^{-}}=-1.61\,[\mu_N]$, and for $J_3=1/2$ states $\mu_{\Delta^{++}}=1.18\,[\mu_N]$, $\mu_{\Delta^{+}}=0.64\,[\mu_N]$, $\mu_{\Delta^{0}}=0.10\,[\mu_N]$, $\mu_{\Delta^{-}}=-0.45\,[\mu_N]$.  
These analyses are consistent with Refs.~\cite{Aubin:2008qp,Slaughter:2011xs}. 

The electric charge density for $\Delta$ isobar states are defined as 
$\rho_E=\frac{1}{2}\rho_{I=0}
+ \<\Psi|I_3|\Psi\>\rho_{I=1}$, 
where $\rho_{I=0}\equiv\left(j_B^0|_{eB\to0}
\right)=\frac{\partial\hat{\mathscr{L}}_{\rm WZW}}{\partial(e{\mathcal{V}_B}_0)}|_{{\mathcal{V}_B}_0\to0, eB\to0}$, $\rho_{I=1}\equiv\left(\frac{1}{3}\sum_{a=1,2,3}\frac{ \Lambda_a}{\<\Psi|\int dV \Lambda_a|\Psi\>}\right)$ and $\Lambda_a\equiv\frac{\d^2 \hat{\mathscr{L}}}{\d \omega_a^2}$. 
The $\Delta$ isobar mean square (MS) electric charge radius is defined as $\<R_\Delta^2\>_E\equiv\< \Psi|\int dV R^2 \rho_E |\Psi\>$, where $R\equiv\sqrt{x^2+y^2+z^2}$.

\begin{figure}[htb]
\centering
\includegraphics[scale=0.46]{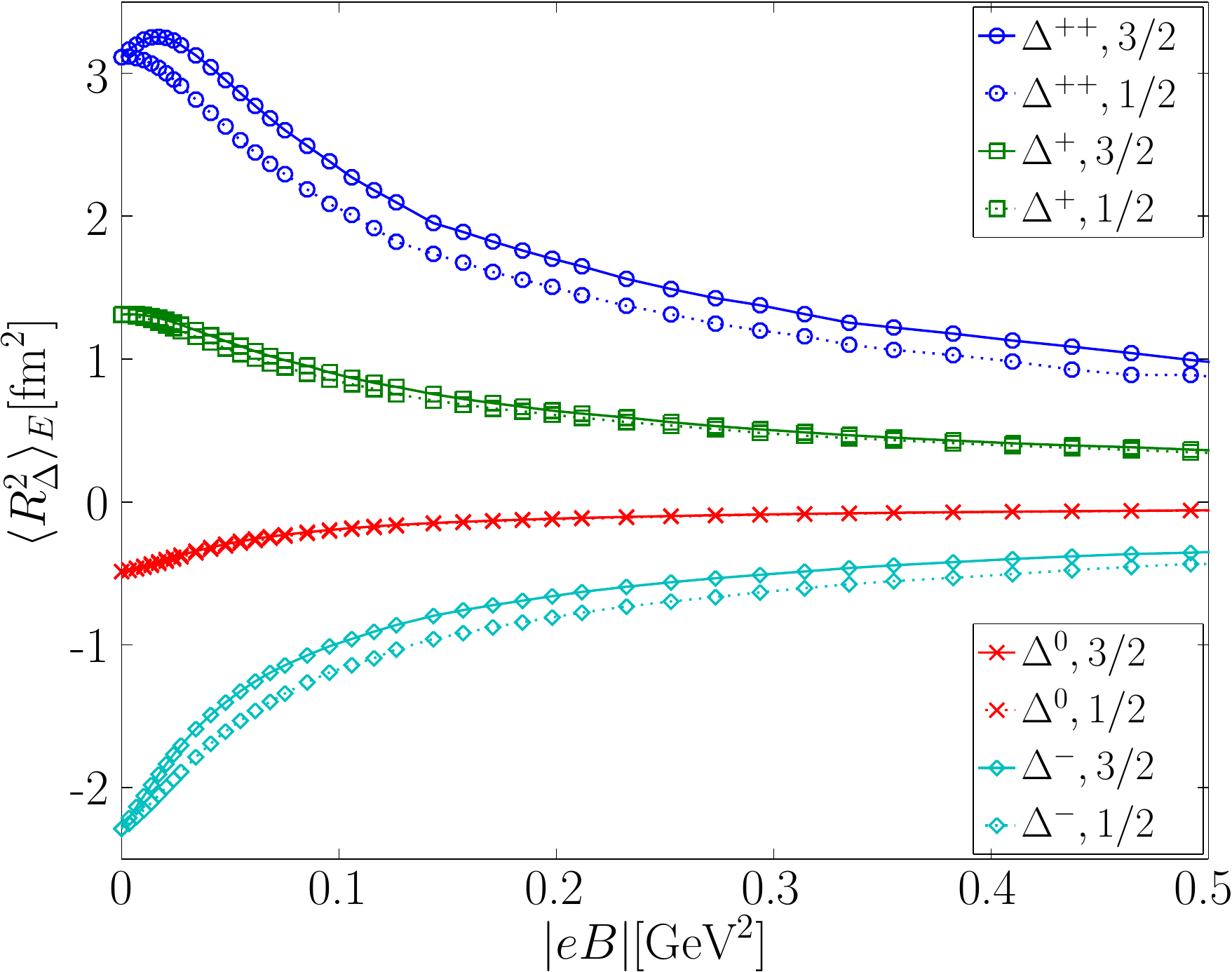} 
\caption{$|e B|$ dependence of the $\Delta$ isobar MS electric charge radius $\<R_\Delta^2\>_E$.
}
\label{fig:bmag_rmsp}
\end{figure}

The $|eB|$ dependence of $\Delta$ isobars MS electric charge radii are shown in Fig.~\ref{fig:bmag_rmsp}. 
Fig.~\ref{fig:bmag_rmsp} shows that for all range of the magnetic field strength, $\<R_\Delta^2\>_E$ of $\Delta^{++}$ and $\Delta^{+}$ are always above zero, whereas $\Delta^{0}$ and $\Delta^{-}$ are always below zero. 
Since the electric charge of $\Delta$ isobars are different, the MS electric charge radii should be different. If $\Delta$ isobars do not have internal structures, the MS electric charge radii of $\Delta$ isobars should satisfy $\Delta^{++}:\Delta^{+}:\Delta^{0}:\Delta^{-}=2:1:0:-1$. 
However, the electric charge of $\Delta$ baryon is constructed by two parts, one part is related to the baryon number current density $\rho_{I=0}$ and another part is related to the iso-vector current density $\rho_{I=1}$. 
For all range of the magnetic field strength, the distribution of $\rho_{I=1}$ is more apart from the central point of the soliton than that of $\rho_{I=0}$, which leads MS electric charge radii of $\Delta^{++}$ and $\Delta^{+}$ are always above zero, whereas $\Delta^{0}$ and $\Delta^{-}$ are always below zero. 
For examples, when the magnetic field is extremely weak, i.e. $|eB|\sim 0$, $\< \Delta|\int dV R^2 \rho_{I=0} |\Delta\>=0.826{\rm \, [fm^2]}$ and $\< \Delta|\int dV R^2 \rho_{I=1} |\Delta\>=1.8{\rm \, [fm^2]}$, thus the MS electric charge radii of $\Delta^{++}$, $\Delta^{+}$, $\Delta^{0}$ and $\Delta^{-}$ are $3.113{\rm \, [fm^2]}$, $1.313{\rm \, [fm^2]}$, $-0.487{\rm \, [fm^2]}$ and $-2.287{\rm \, [fm^2]}$, respectively; 
when the magnetic field is extremely strong, $\< \Delta|\int dV R^2 \rho_{I=0} |\Delta\> \lesssim \< \Delta|\int dV R^2 \rho_{I=1} |\Delta\>$, therefore the MS electric charge radii of $\Delta$ isobars satisfies $\Delta^{++}:\Delta^{+}:\Delta^{0}:\Delta^{-}\simeq2:1:0:-1$.  
Fig.~\ref{fig:bmag_rmsp} also shows that for a nonzero magnetic field region, the magnitude of the MS electric charge radii of $\Delta^{++}$, $\Delta^{+}$ and $\Delta^{0}$ states first increases and then decreases.  This fact is understandable from that: when the magnetic field is weak, there mass decrease, therefore their size will increase; when the magnetic field is strong, the restriction force of charged meson $\pi^{+,-}$ in the $x-y$ plane increase, therefore their size will decrease. 
The magnitude of $\Delta^{-}$ MS electric charge radii always decrease with the increasing of $|eB|$, the reason is that: for a nonzero magnetic field, the $\Delta^{-}$ mass always increases, therefore the size will decrease.

\emph{Conclusions and discussions. ---}
In this letter, the mass, magnetic moment and MS  electric charge radius of $\Delta$ isobar states in a uniform magnetic field were studied in the semi-classical quantization approach of Skyrme model. 

In the vacuum, i.e. $|eB|\sim 0$, it was shown that, the magnetic moment of $\Delta$ isobars can be rewritten by the magnetic moment of proton and neutron. 

In a nonzero magnetic field region, with increasing the strength of magnetic field, it was found that:
\begin{enumerate}[(i)]
\item 
The mass of $\Delta^{++}$, $\Delta^{+}$ and $\Delta^{0}$ states first decreases and then increases, whereas the mass of $\Delta^{-}$ state always increases. 
The minimal mass of $\Delta^{++},J_3=3/2$ is about $1147{\rm \,[MeV]}$ when $|eB|\sim 3.2 m_{\pi}^2$. 

\item 
The magnitude of $\Delta^{++}$, $\Delta^{+}$ and $\Delta^{0}$ magnetic moments first decrease and then increase, whereas the magnitude of $\Delta^{-}$ magnetic moment first increases and then decreases. 

\item
The magnitude of MS electric charge radii corresponding to $\Delta^{++}$, $\Delta^{+}$ and $\Delta^{0}$ states first increase and then decrease, whereas the magnitude of MS electric charge radii corresponding to $\Delta^{-}$ state always decrease.
\end{enumerate}

\begin{table}[hbt]
\centering
\caption{The density of $\Delta$ isobars when $|eB| \sim 10^{-2}{\rm \,[ GeV^2}]$ compared to that in vacuum}
\label{tab:rho_deltas}
\begin{tabular}{ p{1.5cm} p{2cm} p{1.5cm} }
\hline\hline
 & $J_3=3/2$ & $J_3=1/2$\\
\hline
$\rho_{\Delta^{++}}$ & $-8.5\%$ & $-0.1\%$ \\
$\rho_{\Delta^{+}}$ & $-0.9\%$ & $+2.4\%$ \\
$\rho_{\Delta^{0}}$ & $-7.4\%$ & $-5.6\%$ \\
$\rho_{\Delta^{-}}$ & $+19.7\%$ & $+9.8\%$ \\
\hline\hline
\end{tabular}
\end{table}

Science both the mass and size of $\Delta$ baryons depend on the strength of magnetic field, the density of $\Delta$ baryons in the core part of the magnetar ($|eB|\sim10^{-2}{\rm \,[ GeV^2}]$) can be estimated easily, which is given in Table~\ref{tab:rho_deltas}. 
Table~\ref{tab:rho_deltas} shows that, the $\Delta^{++},J_3=3/2$ density decreases about $8.5\%$ and the $\Delta^{-},J_3=3/2$ density increases about $19.7\%$ compared to that in vacuum, respectively. 
Therefore, the equation of state of $\Delta$ baryon media is modified by the magnetic field, which also affects the mass and size limit of magnetar.

This work was supported in part by the National Natural Science Foundation of China (Grant No. 11705094), Natural Science Foundation of Jiangsu Province, China (Grant No. BK20171027), Natural Science Foundation of the Higher Education Institutions of Jiangsu Province, China (Grant No. 17KJB140011), and by the Research Start-up Funding (B.R.~He) of Nanjing Normal University.

\end{document}